\documentstyle[12pt]{article}

\newcommand {\nn}    {\nonumber}
\newcommand {\vs}[1]  { \vspace*{#1 cm} }
\newcounter{eq}
\newcounter{sc}

\newcommand {\MPL}  {Mod.Phys.Lett.}
\newcommand {\NP}   {Nucl.Phys.}
\newcommand {\PL}   {Phys.Lett.}
\newcommand {\PR}   {Phys.Rev.}
\newcommand {\PRL}   {Phys.Rev.Lett.}

\def\overleftrightarrow#1{\vbox{\ialign{##\crcr
 $\leftrightarrow$\crcr\noalign{\kern-1pt\nointerlineskip}
 $\hfil\displaystyle{#1}\hfil$\crcr}}}

\newlength{\minitwocolumn}
\begin{document}

%%%%%%%%%%%%%%%%%%%%%%%%%%%%%%%%%%%%%%%%%%%%%%%%%%%%%%%%%%%%%%%%%%
%%%%%%%%%%%%%%%%%%%%%%%% Title %%%%%%%%%%%%%%%%%%%%%%%%%%%%%%%%%%%
%%%%%%%%%%%%%%%%%%%%%%%%%%%%%%%%%%%%%%%%%%%%%%%%%%%%%%%%%%%%%%%%%%

\begin{flushright}
EDO-EP-23\\
October, 1998\\
\end{flushright}
\vspace{30pt}

%\magnification=\magstep1
\pagestyle{empty}
\baselineskip15pt
%\font\cmssB=cmss17
%\font\cmssS=cmss10

\begin{center}
{\large\bf $SL(2,Z)$ Self-duality of Super D3-bane Action on 
$AdS_5 \times S^5$
 \vskip 1mm
}

\vspace{20mm}

Ichiro Oda
          \footnote{
          E-mail address:\ ioda@edogawa-u.ac.jp
                  }
\\
\vspace{10mm}
          Edogawa University,
          474 Komaki, Nagareyama City, Chiba 270-0198, JAPAN \\

\end{center}

%\maketitle

\vspace{15mm}
\begin{abstract}
It is shown that a supersymmetric and $\kappa$-symmetric D3-bane 
action on $AdS_5 \times S^5$ is mapped into itself by
a duality transformaion, thereby verifying the $SL(2,Z)$
invariance of the D3-brane action in the $AdS_5 \times S^5$ background
as in the flat background.
To this end, we fix the $\kappa$-symmetry
in a gauge which simplifies the classical action in order
to perform an $SO(2)$ rotation of the $N=2$ spinor
index in a manifest way, though this may not be necessary. 
This situation is the same as the case of a super D-string
on $AdS_5 \times S^5$ where it was shown that the super D-string 
action is transformed to a form of the IIB Green-Schwarz superstring 
action with the $SL(2,Z)$ covariant tension in the $AdS_5 \times S^5$ 
background through a duality transformation.
These results strongly suggest that various duality relations originally
found in the flat background may be independent of background geometry,
in other words, the duality transformations in string and p-brane theories
may exist even in general curved space-time.
\vspace{15mm}

\end{abstract}

\newpage
\pagestyle{plain}
\pagenumbering{arabic}
%\setcounter{page}{1}

%%%%%%%%%%%%%%%%%%%%%%%%%%%%%%%%%%%%%%%%%%%%%%%%%%%%%%%%%%%%%%%%%%
%%%%%%%%%%%%%%%%%%%%%%%% Article %%%%%%%%%%%%%%%%%%%%%%%%%%%%%%%%%
%%%%%%%%%%%%%%%%%%%%%%%%%%%%%%%%%%%%%%%%%%%%%%%%%%%%%%%%%%%%%%%%%%

\rm
%%%%%%%%%%%%%%%%%%%%%%%%%%%%%%%%%%%%%%%%%%%%%%%%%%%%%%%%%%%%%%%%%%%%%
%%%%%%%%%%%%%%%%%%%%%%%%%%%%%%   SEC  1    %%%%%%%%%%%%%%%%%%%%%%%%%%
%%%%%%%%%%%%%%%%%%%%%%%%%%%%%%%%%%%%%%%%%%%%%%%%%%%%%%%%%%%%%%%%%%%%%
\section{Introduction}

Recently, studies of string and p-branes in the $AdS_5 \times S^5$ 
background have attracted a great deal of attention.
These studies have been triggered by an interesting conjecture that 
the four dimensional large $N$ super Yang-Mills theory is in a sense 
dual to the type IIB superstring theory on $AdS_5 \times S^5$
\cite{Maldacena}. 
To fully understand this conjecture it is of great importance to 
understand various properties of string and p-brane theories in this 
background in more detail. In particular, $AdS_5 \times S^5$ is 
maximally supersymmetric 'vacuum' with 32 supercharges in addition
to the flat ten dimensional space-time so superstring theory on this
background would give rise to useful informations about the Kaluza-Klein
compactification, quantum corrections through the no-go theorems from
supersymmetry and a background independent formulation of the matrix 
model \cite{Oda3} e.t.c.

A first step towards the construction of supersymmetric string and
D3-brane actions on $AdS_5 \times S^5$ was taken by Metsaev and Tseytlin 
\cite{Metsaev1, Metsaev2}, whose approach is based on the group manifold
method \cite{Fre0}. Their fundamental strategy is to start with the supergroup
$SU(2,2|4)$ directly, consider the coset superspace $SU(2,2|4)/(SO(4,1)
\otimes SO(5))$, solve the Maurer-Cartan equations implied by the
$su(2,2|4)$ superalgebra \cite{Kallosh2}, and then find the supersymmetric
and $\kappa$-symmetric action. Later, the gauge fixing of the
$\kappa$-symmetry
of the action was considered in terms of two different approaches
\cite{Fre, Pesando1, Renata, Kallosh1, Kallosh & Tseytlin}, and
the obtained actions were shown to be actually equivalent to each other
\cite{Pesando2}.

Based on these studies, we have recently investigated the $SL(2,Z)$
S-duality \cite{Schwarz} of type IIB superstring theory on $AdS_5 \times S^5$,
where we have succeeded in proving that the super D-string action 
is also transformed to the type IIB Green-Schwarz superstring action 
\cite{GS} with the $SL(2,Z)$ covariant tension in the $AdS_5 \times S^5$
background in a quantum-mechanically exact manner as in the case of the 
flat Minkowskian background \cite{Oda4}. 
To address this issue one has taken account of
the gauge fixing of the $\kappa$-symmetry since in the classical action
there are terms of higher orders in the spinor variables $\Theta$ 
reflecting the curved nature of the background metric, which seems to make 
it difficult to carry out the $SO(2)$ rotation in the duality 
transformation. This is to be contrasted with the situation in the
flat background where only simple bilinear forms of the spinor
variables appear in the classical action so that we do not have to fix the
gauge symmetries at all. Here it is worth noting that the gauge
fixing procedure does not change the physical contents of a theory
so taking the proper gauge conditions is nothing but a recipe for 
understanding the duality transformation in a manifest way.

In this article, we wish to consider an extension of our previous
studies of a super D-string action on $AdS_5 \times S^5$ \cite{Oda4} 
to a super D3-brane action on the same background \cite{Metsaev2}. 
It is now well-known that the super D3-brane action
in the flat background \cite{Cederwall, Aganagic, Aganagic2, Bergshoeff}
is mapped into an equivalent D3-brane action by the duality
transformation, in this sense, the super D3-brane is self-dual under
$SL(2,Z)$ S-duality \cite{Aganagic2}, 
so that it is natural to ask whether this peculiar feature of the super
D3-brane on the flat background is also inherited to the case of the 
$AdS_5 \times S^5$ background or not. 
We will see that this is indeed the case by using a classical approach.
The validity of the $SL(2,Z)$ duality of string and D3-brane theories
in the $AdS_5 \times S^5$ background may lead to a strong expectation 
that various duality transformations found originally in the flat 
background would hold true even in an arbitrary curved background. We
think that this observation is quite important for future developments
of superstring theory and M-theory since these theories should be
essentially formulated on the basis of not the flat Minkowskian but
the general curved background in a background independent manner
\cite{Oda3}.

The contents of this article are organized as follows. 
In Section 2 we shall review the super D3-brane action in the
$AdS_5 \times S^5$ background which was constructed in  
\cite{Metsaev2}. In Section 3 it will be shown that the
super D3-brane action on this background is actually self-dual
under the $SL(2,Z)$ duality transformation in terms of a classical method. 
To accomplish an appropriate $SO(2)$ rotation in an obvious way will 
require the gauge fixing of the $\kappa$-symmetry. 
The final section will be devoted to
discussions, particularly a summary of the work at hand, the validity of 
the classical approximation and future works.

%%%%%%%%%%%%%%%%%%%%%%%%%%%%%%%%%%%%%%%%%%%%%%%%%%%%%%%%%%%%%%%%%%%%%
%%%%%%%%%%%%%%%%%%%%%%%%%%%%%%   SEC  2    %%%%%%%%%%%%%%%%%%%%%%%%%%
%%%%%%%%%%%%%%%%%%%%%%%%%%%%%%%%%%%%%%%%%%%%%%%%%%%%%%%%%%%%%%%%%%%%%
\section{ Super D3-brane action on $AdS_5 \times S^5$}

We start by not only reviewing the studies of a super D3-brane
action on $AdS_5 \times S^5$ \cite{Metsaev2} but also exposing
a technical detail of the formalism to some extent. 
The $\kappa$-symmetric and reparametrization invariant super D3-brane 
action in the $AdS_5 \times S^5$ background constructed by
Metsaev and Tseytlin \cite{Metsaev2} is given by
%**   2.1 %%%%%%%%%%%%%%%%%%%%%%%%%%%%%%%%%%%%%%%%%%%%%%%%%%%%%%%%%
\begin{eqnarray}
S = S_{DBI} + S_{WZ},
\label{2.1}
\end{eqnarray}
%%%%%%%%%%%%%%%%%%%%%%%%%%%%%%%%%%%%%%%%%%%%%%%%%%%%%%%%%%%%%%%%%%%
with
%**   2.2 %%%%%%%%%%%%%%%%%%%%%%%%%%%%%%%%%%%%%%%%%%%%%%%%%%%%%%%%%
\begin{eqnarray}
&{}& S_{DBI} = -  \int_{M_4} d^4 \sigma 
\sqrt{- \det ( G_{ij} + {\cal F}_{ij} )}, \nn\\
&{}& S_{WZ} =  \int_{M_5} H_5 = \int_{M_4 = \partial M_5} 
\Omega_4, \nn\\
&{}&  H_5  = d \Omega_4,
\label{2.2}
\end{eqnarray}
%%%%%%%%%%%%%%%%%%%%%%%%%%%%%%%%%%%%%%%%%%%%%%%%%%%%%%%%%%%%%%%%%%%
where the 3-brane tension is chosen to be a unity and 
the 5-form $H_5$ in the Wess-Zumino term is given by
%**   2.3 %%%%%%%%%%%%%%%%%%%%%%%%%%%%%%%%%%%%%%%%%%%%%%%%%%%%%%%%%
\begin{eqnarray}
H_5 &=& i \bar{L} \wedge (\frac{1}{6} \hat{L} \wedge \hat{L} \wedge 
\hat{L} {\cal E} + {\cal F} \wedge \hat{L} {\cal I}) \wedge L \nn\\
& & {} + \frac{1}{30} \epsilon^{a_1 \cdots a_5} L^{a_1} \wedge \cdots
\wedge L^{a_5} + \frac{1}{30} \epsilon^{a'_1 \cdots a'_5} L^{a'_1} 
\wedge \cdots \wedge L^{a'_5},
\label{2.3}
\end{eqnarray}
%%%%%%%%%%%%%%%%%%%%%%%%%%%%%%%%%%%%%%%%%%%%%%%%%%%%%%%%%%%%%%%%%%%
where $i$ and $j$ run over the world-volume indices
0, 1, 2 and 3, and we have defined 
%**   2.4 %%%%%%%%%%%%%%%%%%%%%%%%%%%%%%%%%%%%%%%%%%%%%%%%%%%%%%%%%
\begin{eqnarray}
&{}& G_{ij} = L^{\hat{a}}_i L^{\hat{a}}_j, \ 
\hat{L} \equiv L^{\hat{a}} \Gamma^{\hat{a}}, \nn\\
&{}& {\cal E} = i \sigma_2 = \pmatrix{
0  & 1 \cr -1 & 0 \cr }, \ {\cal I} = \sigma_1 = \pmatrix{
0  & 1 \cr 1 & 0 \cr },  \ {\cal K} = \sigma_3 = \pmatrix{
1  & 0 \cr 0 & -1 \cr }.
\label{2.4}
\end{eqnarray}
%%%%%%%%%%%%%%%%%%%%%%%%%%%%%%%%%%%%%%%%%%%%%%%%%%%%%%%%%%%%%%%%%%%
Here $L^I$ and $L^{\hat{a}}$ are the Cartan 1-form spinor superfield
and the vector superfield, respectively. They satisfy the Maurer-Cartan
equations for $su(2, 2|4)$ superalgebra \cite{Metsaev1, Metsaev2}
%**   2.5 %%%%%%%%%%%%%%%%%%%%%%%%%%%%%%%%%%%%%%%%%%%%%%%%%%%%%%%%%
\begin{eqnarray}
d L^{\hat {a}} &=& - L^{\hat{a}\hat{b}} \wedge L^{\hat {b}}
- i \bar{L} \Gamma^{\hat{a}} \wedge L, \nn\\ 
d L &=& \frac{i}{2} \sigma_+ {\hat{L}} \wedge {\cal E} L
-\frac{1}{4} L^{\hat{a}\hat{b}} \Gamma^{\hat{a}\hat{b}} \wedge L, \nn\\
d \bar{L} &=& \frac{i}{2} \bar{L} {\cal E} \wedge {\hat{L}} \sigma_+ 
-\frac{1}{4} \bar{L} \Gamma^{\hat{a}\hat{b}} \wedge L^{\hat{a}\hat{b}},
\label{2.5}
\end{eqnarray}
%%%%%%%%%%%%%%%%%%%%%%%%%%%%%%%%%%%%%%%%%%%%%%%%%%%%%%%%%%%%%%%%%%%
and $\sigma_i$ are the Pauli matrices, which operate on $N=2$
spinor indices $I, J = 1, 2$. 
Throughout this article we follow the conventions and notations of 
references \cite{Metsaev1, Metsaev2, Kallosh & Tseytlin}. 
 
In the explicit parametrization $G(X, \Theta) = g(X) e^{\Theta Q}$
where $X^{\hat{m}}$, $\Theta^I$, and $Q_I$ are respectively the 
bosonic and fermionic space-time coordinates and 32-component
supercharges \cite{Metsaev1}, the action (\ref{2.1}) takes the form 
%**   2.6 %%%%%%%%%%%%%%%%%%%%%%%%%%%%%%%%%%%%%%%%%%%%%%%%%%%%%%%%%
\begin{eqnarray}
S &=& -  \int_{M_4} d^4 \sigma 
\sqrt{- \det ( G_{ij} + {\cal F}_{ij} )}
+ 2i \int_{M_4} \int^1_0 ds \ \bar{\Theta} (\frac{1}{6} \hat{L}_s
\wedge \hat{L}_s \wedge \hat{L}_s {\cal E} 
+ {\cal F}_s \wedge \hat{L}_s {\cal I}) \wedge L_s \nn\\
& & {} + \int_{M_5} F_5,
\label{2.6}
\end{eqnarray}
%%%%%%%%%%%%%%%%%%%%%%%%%%%%%%%%%%%%%%%%%%%%%%%%%%%%%%%%%%%%%%%%%%%
where 
%**   2.7 %%%%%%%%%%%%%%%%%%%%%%%%%%%%%%%%%%%%%%%%%%%%%%%%%%%%%%%%%
\begin{eqnarray}
&{}& {\cal F} = F + 2i \int^1_0 ds \bar{\Theta} \hat{L}_{s}
\wedge {\cal K} L_{s}, \nn\\
&{}& {\cal F}_s = F + 2i \int^s_0 ds' \bar{\Theta} \hat{L}_{s'}
\wedge {\cal K} L_{s'}, \nn\\
&{}& F_5 = \frac{1}{30} (\epsilon^{a_1 \cdots a_5} e^{a_1} \wedge
\cdots \wedge e^{a_5} + \epsilon^{a'_1 \cdots a'_5} e^{a'_1} \wedge
\cdots \wedge e^{a'_5}),
\label{2.7}
\end{eqnarray}
%%%%%%%%%%%%%%%%%%%%%%%%%%%%%%%%%%%%%%%%%%%%%%%%%%%%%%%%%%%%%%%%%%%
with $F = dA$ being the $U(1)$ gauge potential and 
$L^{\hat{a}}|_{\Theta=0}
\equiv e^{\hat{a}}$ (the vielbein 1-form). We have used the standard 
rescaling trick $\Theta \rightarrow s \Theta$ and the Maurer-Cartan
equations in deriving the second and third terms in the right-hand 
side of Eq.(\ref{2.6}). 
Note that the two terms in $\int_{M_5} F_5$ describes that the 
self-dual RR 5-form field strength takes nontrivial values on $AdS_5$ 
and $S^5$, respectively. 
(Here we adopt the convention that the radii of $AdS_5$ and $S^5$
are a unity.)
Actually this coupling precisely corresponds to the Freund-Rubin 
spontaneous 
compactification mechanism \cite{Freund} in ten dimensions. 
It is surprising that 
the requirement of supersymmetry and $\kappa$-symmetry naturally leads
to this coupling. 
The Cartan invariant 1-forms $L^I = L^I_{s=1}$  
and $L^{\hat{a}} = L^{\hat{a}}_{s=1}$ are determined by solving the 
Maurer-Cartan equations (\ref{2.5}) as follows \cite{Kallosh2} 
%**   2.8 %%%%%%%%%%%%%%%%%%%%%%%%%%%%%%%%%%%%%%%%%%%%%%%%%%%%%%%%%
\begin{eqnarray}
L^I_s &=& \left( \frac{\sinh(s {\cal M})}{\cal M} D\Theta \right)^I, \nn\\
L^{\hat{a}}_s &=& e^{\hat{a}}_{\hat{m}}(X)dX^{\hat{m}}
- 4i {\bar{\Theta}}^I \Gamma^{\hat{a}} \left( \frac{\sinh^2
(\frac{1}{2} s {\cal M})}{{\cal M}^2} D\Theta \right)^I,
\label{2.8}
\end{eqnarray}
%%%%%%%%%%%%%%%%%%%%%%%%%%%%%%%%%%%%%%%%%%%%%%%%%%%%%%%%%%%%%%%%%%%
with 
%**   2.9 %%%%%%%%%%%%%%%%%%%%%%%%%%%%%%%%%%%%%%%%%%%%%%%%%%%%%%%%%
\begin{eqnarray}
({\cal M}^2)^{IL} &=& \epsilon^{IJ} (-\gamma^a \Theta^J
{\bar{\Theta}}^L \gamma^a +  \gamma^{a'} \Theta^J
{\bar{\Theta}}^L \gamma^{a'})
+ \frac{1}{2} \epsilon^{KL} (\gamma^{ab} \Theta^I
{\bar{\Theta}}^K \gamma^{ab} -  \gamma^{a'b'} \Theta^I
{\bar{\Theta}}^K \gamma^{a'b'}), \nn\\
(D\Theta)^I &=& \left[d + \frac{1}{4}(\omega^{ab}\gamma_{ab}
+ \omega^{a'b'}\gamma_{a'b'}) \right] \Theta^I
- \frac{1}{2} i \epsilon^{IJ}(e^a \gamma_a + i e^{a'}
\gamma_{a'}) \Theta^J.
\label{2.9}
\end{eqnarray}
%%%%%%%%%%%%%%%%%%%%%%%%%%%%%%%%%%%%%%%%%%%%%%%%%%%%%%%%%%%%%%%%%%%

Now let us present the $\kappa$-transformation of
the super D3-brane action (\ref{2.1}) whose concrete expressions
are given by \cite{Metsaev2} 
%**   2.10 %%%%%%%%%%%%%%%%%%%%%%%%%%%%%%%%%%%%%%%%%%%%%%%%%%%%%%%%%
\begin{eqnarray}
\delta_{\kappa} \Theta^I = \kappa^I,
\label{2.10}
\end{eqnarray}
%%%%%%%%%%%%%%%%%%%%%%%%%%%%%%%%%%%%%%%%%%%%%%%%%%%%%%%%%%%%%%%%%%%
and the projection $\Gamma$ is \cite{Cederwall}
%**   2.11 %%%%%%%%%%%%%%%%%%%%%%%%%%%%%%%%%%%%%%%%%%%%%%%%%%%%%%%%%
\begin{eqnarray}
\Gamma \kappa  &=& \kappa, \ \Gamma^2 = 1, \ Tr\Gamma = 0, \nn\\
\Gamma &=& \frac{\epsilon^{i_1 \cdots i_4}}{\sqrt{- \det ( G_{ij} + 
{\cal F}_{ij} )}} (\frac{1}{4!} \Gamma_{i_1 \cdots i_4} {\cal E} 
- \frac{1}{4} \Gamma_{i_1 i_2} {\cal F}_{i_3 i_4} {\cal I}
+ \frac{1}{8} {\cal F}_{i_1 i_2} {\cal F}_{i_3 i_4} {\cal E}),
\label{2.11}
\end{eqnarray}
%%%%%%%%%%%%%%%%%%%%%%%%%%%%%%%%%%%%%%%%%%%%%%%%%%%%%%%%%%%%%%%%%%%
with the definition of $\Gamma_{i_1 \cdots i_n} \equiv 
{\hat{L}}_{[i_1} \cdots {\hat{L}}_{i_n]}$. Moreover, various 
superfields must transform under the $\kappa$-transformation
as follows \cite{Metsaev1, Metsaev2}
%**   2.12 %%%%%%%%%%%%%%%%%%%%%%%%%%%%%%%%%%%%%%%%%%%%%%%%%%%%%%%%%
\begin{eqnarray}
\delta_{\kappa} L^{\hat {a}} &=& 2i \bar{L} \Gamma^{\hat{a}} 
\delta_{\kappa} \Theta, \nn\\
\delta_{\kappa} L &=& d \delta_{\kappa} \Theta - \frac{i}{2} \sigma_+
{\hat{L}} {\cal E} \delta_{\kappa} \Theta + \frac{1}{4}
L^{\hat{a}\hat{b}} \Gamma^{\hat{a}\hat{b}} \delta_{\kappa} \Theta, \nn\\
\delta_{\kappa} \bar{L} &=& d \delta_{\kappa} \bar{\Theta} +
\frac{i}{2} \delta_{\kappa} \bar{\Theta} {\cal E} {\hat{L}}\sigma_+
- \frac{1}{4} \delta_{\kappa} \bar{\Theta} \Gamma^{\hat{a}\hat{b}} 
L^{\hat{a}\hat{b}}, \nn\\
\delta_{\kappa} G_{ij} &=& 2i (\bar{L}_i {\hat{L}}_j + 
\bar{L}_j {\hat{L}}_i) \delta_{\kappa} \Theta, \nn\\
\delta_{\kappa} {\cal F}_{ij} &=& 2i (\bar{L}_i {\cal K} {\hat{L}}_j 
- \bar{L}_j {\cal K} {\hat{L}}_i) \delta_{\kappa} \Theta.
\label{2.12}
\end{eqnarray}
%%%%%%%%%%%%%%%%%%%%%%%%%%%%%%%%%%%%%%%%%%%%%%%%%%%%%%%%%%%%%%%%%%%

Then it is a little tedious but straightforward to derive
%**   2.13 %%%%%%%%%%%%%%%%%%%%%%%%%%%%%%%%%%%%%%%%%%%%%%%%%%%%%%%%%
\begin{eqnarray}
\delta_{\kappa} H_5 = d \Lambda_4,
\label{2.13}
\end{eqnarray}
%%%%%%%%%%%%%%%%%%%%%%%%%%%%%%%%%%%%%%%%%%%%%%%%%%%%%%%%%%%%%%%%%%%
where
%**   2.14 %%%%%%%%%%%%%%%%%%%%%%%%%%%%%%%%%%%%%%%%%%%%%%%%%%%%%%%%%
\begin{eqnarray}
\Lambda_4 = 2i \bar{L} \wedge (\frac{1}{6} \hat{L} \wedge \hat{L} \wedge 
\hat{L} {\cal E} + {\cal F} \wedge \hat{L} {\cal I}) 
\delta_{\kappa} \Theta.
\label{2.14}
\end{eqnarray}
%%%%%%%%%%%%%%%%%%%%%%%%%%%%%%%%%%%%%%%%%%%%%%%%%%%%%%%%%%%%%%%%%%%

In order to prove the equation (\ref{2.13}), we have made use of
the Maurer-Cartan equations (\ref{2.5}) and the
following cyclic Fierz identity for fermionic spinors $A, B, C, D$
\cite{Cederwall}
%**   2.15 %%%%%%%%%%%%%%%%%%%%%%%%%%%%%%%%%%%%%%%%%%%%%%%%%%%%%%%%%
\begin{eqnarray}
(\bar{A} \Gamma^{\hat{a}} B) \cdot (\bar{C} \Gamma^{\hat{a}} D)
= -\frac{1}{2} \left[(\bar{A} \Gamma^{\hat{a}} e_I D) \cdot
(\bar{B} \Gamma^{\hat{a}} e_I C) - (\bar{A} \Gamma^{\hat{a}} e_I C) 
\cdot (\bar{B} \Gamma^{\hat{a}} e_I D) \right],
\label{2.15}
\end{eqnarray}
%%%%%%%%%\%%%%%%%%%%%%%%%%%%%%%%%%%%%%%%%%%%%%%%%%%%%%%%%%%%%%%%%%%%
where $e_I = \{1, {\cal E}, {\cal I}, {\cal K}\}$. 
Then after a little tedious calculations it turns out that the super
D3-brane action on $AdS_5 \times S^5$ is invariant under 
the $\kappa$ transformation
%**   2.16 %%%%%%%%%%%%%%%%%%%%%%%%%%%%%%%%%%%%%%%%%%%%%%%%%%%%%%%%%
\begin{eqnarray}
\delta_{\Gamma\kappa} S_{DBI} + \delta_{\kappa} S_{WZ} = 0,
\label{2.16}
\end{eqnarray}
%%%%%%%%%%%%%%%%%%%%%%%%%%%%%%%%%%%%%%%%%%%%%%%%%%%%%%%%%%%%%%%%%%%
where we have used various matrix formulas and $\Gamma$ matrix
identity such as
%**   2.17 %%%%%%%%%%%%%%%%%%%%%%%%%%%%%%%%%%%%%%%%%%%%%%%%%%%%%%%%%
\begin{eqnarray}
\delta \det M &=& \det M \cdot Tr(M^{-1} \delta M), \nn\\
(G + {\cal F})^{-1 ji} &=& (G - {\cal F})^{-1 ij}, \nn\\
(1 \pm G^{ik} {\cal F}_{kj})^{-1} &=& (1 + (G^{ik} {\cal F}_{kj})^2)^{-1}
(1 \mp G^{ik} {\cal F}_{kj}), \nn\\
\Gamma^{\hat{a}} \Gamma^{\hat{a_1} \cdots \hat{a_{2n}}}
&=& \Gamma^{\hat{a} \hat{a}_1 \cdots \hat{a}_{2n}}
+ 2 n \eta^{\hat{a} [\hat{a}_1} \Gamma^{\hat{a}_2 \cdots \hat{a}_{2n}]}.
\label{2.17}
\end{eqnarray}
%%%%%%%%%%%%%%%%%%%%%%%%%%%%%%%%%%%%%%%%%%%%%%%%%%%%%%%%%%%%%%%%%%%

%%%%%%%%%%%%%%%%%%%%%%%%%%%%%%%%%%%%%%%%%%%%%%%%%%%%%%%%%%%%%%%%%%%%%
%%%%%%%%%%%%%%%%%%%%%%%%%%%%%%   SEC  3    %%%%%%%%%%%%%%%%%%%%%%%%%%
%%%%%%%%%%%%%%%%%%%%%%%%%%%%%%%%%%%%%%%%%%%%%%%%%%%%%%%%%%%%%%%%%%%%%
\section{ SL(2,Z) self-duality}

We now turn our attention to a proof of self-duality of the super
D3-brane action under the $SL(2,Z)$ duality transformation. 
In the flat background this issue was studied in \cite{Tseytlin}
for the bosonic case and in \cite{Aganagic2} for the supersymmetric
case.
One of the crucial issues left so far unsolved is to examine
whether this self-duality is valid even in a curved background.
In fact, in the final section in \cite{Aganagic2}, it is stated
that "..... For the most part, our analysis has been classical
and limited to flat backgrounds. The results should not depend on 
these restrictions, however." Motivated by the sentences,
we have recently clarified that a type IIB superstring action
on $AdS_5 \times S^5$ exhibits the $SL(2,Z)$ S-duality as
in the case of the flat background \cite{Oda4}. In this section,
we wish to show the expected $SL(2,Z)$ invariance of the super
D3-brane reviewed in the previous section.

In the work of the super D-string on $AdS_5 \times S^5$ \cite{Oda4} 
we have used the path integral based on the first-order Hamiltonian 
formalism \cite{de Alwis, Oda1} since this method is effective
in showing the equivalence at the quantum level.
Instead, in the present article, we shall employ a classical
approach developed  in \cite{Tseytlin, Aganagic2}. 
The reason why we use the classical approach will be
argued to some extent in the final section. 

Let us consider the following action:
%**   3.1 %%%%%%%%%%%%%%%%%%%%%%%%%%%%%%%%%%%%%%%%%%%%%%%%%%%%%%%%%
\begin{eqnarray}
S &=& -  \int_{M_4} d^4 \sigma 
\sqrt{- \det ( G_{ij} + {\cal F}_{ij} )}
+ 2i \int_{M_4} \int^1_0 ds \ \bar{\Theta} (\frac{1}{6} \hat{L}_s
\wedge \hat{L}_s \wedge \hat{L}_s {\cal E} 
+ {\cal F}_s \wedge \hat{L}_s {\cal I}) \wedge L_s \nn\\
& & {} + \int_{M_5} F_5 + \int_{M_4} \frac{1}{2} C_0 
F \wedge F,
\label{3.1}
\end{eqnarray}
%%%%%%%%%%%%%%%%%%%%%%%%%%%%%%%%%%%%%%%%%%%%%%%%%%%%%%%%%%%%%%%%%%%
where we have added the topological term $\frac{1}{2} C_0 
F \wedge F$ with a constant axion $C_0$ to the action 
(\ref{2.6}) for later convenience. However, this expression
(\ref{3.1}) is not so illuminating for the purpose of an analysis of
the duality transformation. Thus, instead, we shall consider
an alternative form of the action equivalent
to (\ref{3.1}), which is given by
%**   3.2 %%%%%%%%%%%%%%%%%%%%%%%%%%%%%%%%%%%%%%%%%%%%%%%%%%%%%%%%%
\begin{eqnarray}
S = -  \int_{M_4} d^4 \sigma 
\sqrt{- \det ( G_{ij} + {\cal F}_{ij} )} 
+ \int_{M_4} (C_4 + C_2 \wedge {\cal F} +
\frac{1}{2} C_0 F \wedge F),
\label{3.2}
\end{eqnarray}
%%%%%%%%%%%%%%%%%%%%%%%%%%%%%%%%%%%%%%%%%%%%%%%%%%%%%%%%%%%%%%%%%%%
whose superficial expression is very similar to that adopted in 
\cite{Aganagic2}, but of course the physical contents are different.
However, we will see that this form of the action would be
useful in the analysis of the self-duality of the super D3-brane
action on $AdS_5 \times S^5$ since we can follow similar 
arguments to as in \cite{Aganagic2}.

To make contact with (\ref{3.1}) one has to express the
4-form field $C_4$ and 2-form field $C_2$ in terms of the superfields
and the supercoordinates existing in the theory at hand.
This is provided by the relation
%**   3.3 %%%%%%%%%%%%%%%%%%%%%%%%%%%%%%%%%%%%%%%%%%%%%%%%%%%%%%%%%
\begin{eqnarray}
H_5 = d \Omega_4, \ \Omega_4 = C_4 + C_2 \wedge {\cal F},
\label{3.3}
\end{eqnarray}
%%%%%%%%%%%%%%%%%%%%%%%%%%%%%%%%%%%%%%%%%%%%%%%%%%%%%%%%%%%%%%%%%%%
by which we have the coupled differential equations to be solved 
for $C_4$ and $C_2$
%**   3.4 %%%%%%%%%%%%%%%%%%%%%%%%%%%%%%%%%%%%%%%%%%%%%%%%%%%%%%%%%
\begin{eqnarray}
dC_4 - C_2 \wedge db_2 
&=& i \bar{L} \wedge \frac{1}{6} \hat{L} \wedge \hat{L} \wedge 
\hat{L} {\cal E} \wedge L \nn\\
& & {} + \frac{1}{30} \epsilon^{a_1 \cdots a_5} L^{a_1} \wedge \cdots
\wedge L^{a_5} + \frac{1}{30} \epsilon^{a'_1 \cdots a'_5} L^{a'_1} 
\wedge \cdots \wedge L^{a'_5}, \nn\\
dC_2 \wedge {\cal F} &=& i \bar{L} \wedge {\cal F} \wedge \hat{L} 
{\cal I} \wedge L,
\label{3.4}
\end{eqnarray}
%%%%%%%%%%%%%%%%%%%%%%%%%%%%%%%%%%%%%%%%%%%%%%%%%%%%%%%%%%%%%%%%%%%
where we have defined as 
%**   3.5 %%%%%%%%%%%%%%%%%%%%%%%%%%%%%%%%%%%%%%%%%%%%%%%%%%%%%%%%%
\begin{eqnarray}
{\cal F} &=& F - b_2, \nn\\
b_2 &=& - 2i \int^1_0 ds \bar{\Theta} \hat{L}_{s}
\wedge {\cal K} L_{s}.
\label{3.5}
\end{eqnarray}
%%%%%%%%%%%%%%%%%%%%%%%%%%%%%%%%%%%%%%%%%%%%%%%%%%%%%%%%%%%%%%%%%%%
Then it is relatively straightforward to solve the second equation 
in Eq.(\ref{3.4})
whose result is given by
%**   3.6 %%%%%%%%%%%%%%%%%%%%%%%%%%%%%%%%%%%%%%%%%%%%%%%%%%%%%%%%%
\begin{eqnarray}
C_2 = 2i \int^1_0 ds \bar{\Theta} \hat{L}_{s}
\wedge {\cal I} L_{s},
\label{3.6}
\end{eqnarray}
%%%%%%%%%%%%%%%%%%%%%%%%%%%%%%%%%%%%%%%%%%%%%%%%%%%%%%%%%%%%%%%%%%%
where the Maurer-Cartan equations and the trick $\Theta \rightarrow
s \Theta$ were utilized again.
The first equation in Eq.(\ref{3.4}) can be also solved with the help
of Eq.(\ref{3.6}), but it is unnecessary to use the concrete
expression of $C_4$ for the present consideration 
so we omit to write it down explicitly. 

We are now in a position to think of the duality transformation
of the action (\ref{3.2}). Following a similar path of derivation
to the super D3-brane in the flat background \cite{Aganagic2},
first of all, we have to add $\int d^4 \sigma \frac{1}{2}
\tilde{H}^{ij} (F_{ij} - 2 \partial_i A_j)$ to the action
(\ref{3.2}), and then take the
variation with respect to $A_i$, which gives us the solution
$\tilde{H}^{ij} = \epsilon^{ijkl} \partial_k B_l$ where
$B_i$ is a vector field. After substituting this solution into
the action (\ref{3.2}) and solving the variational equation
for $F_{ij}$ to rewrite the action in terms of $B_i$ 
instead of $F_{ij}$, we arrive at the dual action of (\ref{3.2})
%**   3.7 %%%%%%%%%%%%%%%%%%%%%%%%%%%%%%%%%%%%%%%%%%%%%%%%%%%%%%%%%
\begin{eqnarray}
S_D = -  \int_{M_4} d^4 \sigma 
\sqrt{- \det \left[ G_{ij} + \frac{1}{\sqrt{1+C_0^2}} 
(\tilde{F}_{ij} + C_{ij} + C_0 b_{ij}) \right]} 
+ \int_{M_4} \Omega_D,
\label{3.7}
\end{eqnarray}
%%%%%%%%%%%%%%%%%%%%%%%%%%%%%%%%%%%%%%%%%%%%%%%%%%%%%%%%%%%%%%%%%%%
where $\tilde{F} = dB$ and $\Omega_D$ is defined by
%**   3.8 %%%%%%%%%%%%%%%%%%%%%%%%%%%%%%%%%%%%%%%%%%%%%%%%%%%%%%%%%
\begin{eqnarray}
\Omega_D &=& C_4 - b_2 \wedge C_2 - \frac{1}{2} C_0 b_2 \wedge
b_2 + b_2 \wedge (\tilde{F} + C_2 + C_0 b_2)  \nn\\
& & {} - \frac{C_0}{2(1+C_0^2)} (\tilde{F} + C_2 + C_0 b_2) \wedge
(\tilde{F} + C_2 + C_0 b_2).
\label{3.8}
\end{eqnarray}
%%%%%%%%%%%%%%%%%%%%%%%%%%%%%%%%%%%%%%%%%%%%%%%%%%%%%%%%%%%%%%%%%%%

In order to prove the self-duality one has to show that $S_D$ has 
the same form as the original action (\ref{3.2}). To this aim,
we perform an appropriate $SO(2)$ rotation of the spinor coordinates
$\Theta^I$, which amounts to the following $SO(2)$ rotation of the 
'Pauli matrices' 
%**   3.9 %%%%%%%%%%%%%%%%%%%%%%%%%%%%%%%%%%%%%%%%%%%%%%%%%%%%%%%%%
\begin{eqnarray}
{\cal I'} &=& - \frac{1}{\sqrt{1+C_0^2}} ({\cal K} + 
C_0 {\cal I}), \nn\\
{\cal K'} &=&  \frac{1}{\sqrt{1+C_0^2}} ({\cal I} - 
C_0 {\cal K}).
\label{3.9}
\end{eqnarray}
%%%%%%%%%%%%%%%%%%%%%%%%%%%%%%%%%%%%%%%%%%%%%%%%%%%%%%%%%%%%%%%%%%%
A delicate problem of how one performs the $SO(2)$ rotation of the spinor 
coordinates $\Theta^I$ will discussed below in detail. Then, we
obtain
%**   3.10 %%%%%%%%%%%%%%%%%%%%%%%%%%%%%%%%%%%%%%%%%%%%%%%%%%%%%%%%%
\begin{eqnarray}
\frac{1}{\sqrt{1+C_0^2}} (\tilde{F} + C_2 + C_0 b_2) 
= \tilde{F'} - b'_2 \equiv \tilde{\cal F'},
\label{3.10}
\end{eqnarray}
%%%%%%%%%%%%%%%%%%%%%%%%%%%%%%%%%%%%%%%%%%%%%%%%%%%%%%%%%%%%%%%%%%%
where we have defined $\tilde{F'} \equiv \frac{1}{\sqrt{1+C_0^2}} 
\tilde{F}$.
Consequently, the Dirac-Born-Infeld term in $S_D$ can be rewritten
to be the same form as the original action. 
Moreover, from (\ref{3.4})-(\ref{3.6}) and (\ref{3.8})-(\ref{3.10}),
we can evaluate the Wess-Zumino part as follows
%**   3.11 %%%%%%%%%%%%%%%%%%%%%%%%%%%%%%%%%%%%%%%%%%%%%%%%%%%%%%%%%
\begin{eqnarray}
d\Omega_D &\equiv& dC'_4 - C'_2 \wedge db'_2 + dC'_2  \wedge
\tilde{\cal F'} \nn\\
&=& i \bar{L} \wedge (\frac{1}{6} \hat{L} \wedge \hat{L} \wedge 
\hat{L} {\cal E'} + \tilde{{\cal F'}} \wedge \hat{L} {\cal I'}) 
\wedge L \nn\\
& & {} + \frac{1}{30} \epsilon^{a_1 \cdots a_5} L^{a_1} \wedge \cdots
\wedge L^{a_5} + \frac{1}{30} \epsilon^{a'_1 \cdots a'_5} L^{a'_1} 
\wedge \cdots \wedge L^{a'_5},
\label{3.11}
\end{eqnarray}
%%%%%%%%%%%%%%%%%%%%%%%%%%%%%%%%%%%%%%%%%%%%%%%%%%%%%%%%%%%%%%%%%%%
where we have used the fact that ${\cal E'} = {\cal E}$ under
the $SO(2)$ rotation (\ref{3.9}). Accordingly, the dual action
of the super D3-brane action on $AdS_5 \times S^5$ can be
rewritten as
%**   3.12 %%%%%%%%%%%%%%%%%%%%%%%%%%%%%%%%%%%%%%%%%%%%%%%%%%%%%%%%%
\begin{eqnarray}
S = -  \int_{M_4} d^4 \sigma 
\sqrt{- \det ( G_{ij} + \tilde{{\cal F'}}_{ij} )} 
+ \int_{M_4} (C'_4 + C'_2 \wedge \tilde{{\cal F'}} 
+ \frac{1}{2} C_0 \tilde{F'} \wedge \tilde{F'}),
\label{3.12}
\end{eqnarray}
%%%%%%%%%%%%%%%%%%%%%%%%%%%%%%%%%%%%%%%%%%%%%%%%%%%%%%%%%%%%%%%%%%%
In this way, we can prove the self-duality of the super D3-brane
action on $AdS_5 \times S^5$ in a similar way to the case of
the flat background.

To understand the $SL(2,Z)$ duality more clearly, it is useful
to introduce a constant dilaton background in addition to a
constant axion background. Provided that one starts with
the action \cite{Aganagic2}
%**   3.13 %%%%%%%%%%%%%%%%%%%%%%%%%%%%%%%%%%%%%%%%%%%%%%%%%%%%%%%%%
\begin{eqnarray}
S &=& -  \int_{M_4} d^4 \sigma 
\sqrt{- \det ( G_{ij} + e^{-\frac{\phi}{2}} F_{ij} - b_{ij})} 
+ \int_{M_4} (C_4 + C_2 \wedge ({e^{-\frac{\phi}{2}} F - b_2})
\nn\\
& &{} + \frac{1}{2} C_0 F \wedge F),
\label{3.13}
\end{eqnarray}
%%%%%%%%%%%%%%%%%%%%%%%%%%%%%%%%%%%%%%%%%%%%%%%%%%%%%%%%%%%%%%%%%%%
one can show that this supersymmetric and $\kappa$-symmetric
action is also self-dual with the desired $SL(2,Z)$ transformation
law of the dilaton and the axion
%**   3.14 %%%%%%%%%%%%%%%%%%%%%%%%%%%%%%%%%%%%%%%%%%%%%%%%%%%%%%%%%
\begin{eqnarray}
e^{-\phi} &\rightarrow& \frac{1}{e^{-\phi} + e^{\phi} C_0^2}, \nn\\
C_0 &\rightarrow& - \frac{e^{\phi} C_0}{e^{-\phi} + e^{\phi} C_0^2}.
\label{3.14}
\end{eqnarray}
%%%%%%%%%%%%%%%%%%%%%%%%%%%%%%%%%%%%%%%%%%%%%%%%%%%%%%%%%%%%%%%%%%%

Before closing this section, we should discuss the $SO(2)$ rotation 
of the spinor
coordinates, or equivalently, the $SO(2)$ rotation of the 
'Pauli matrices' (\ref{3.9}).  As already studied in the 
duality transformation of a type IIB superstring on $AdS_5 \times 
S^5$ \cite{Oda4}, at the present case we shall also take a convenient 
gauge condition of the $\kappa$-symmetry to make the fermionic
structure of the classical action simpler. 
This is because owing to a complicated dependence on the 
spinor coordinates $\Theta$ of the classical action as seen in 
Eqs.(\ref{2.8}) and (\ref{2.9}) it is not always obvious that 
we can carry out the desired $SO(2)$ rotation in the classical action.
Of course, we do not intend to exclude other possibilities
of performing the $SO(2)$ rotation at all. For instance, 
as another attempt, it turns out to be advantageous to
parametrize the supergroup element $G(X, \Theta)$ in a
specific parametrization reflecting a underlying geometrical
structure, by which the classical action would become a rather simpler
form than the case at hand. Then there might exist some ingenious 
prescription for the $SO(2)$ rotation of the spinor coordinates
even without fixing any local symmetries. 

In this article, following the previous work \cite{Oda4}
we shall utilize the recently developed quantization
method \cite{Kallosh1, Kallosh & Tseytlin} 
\footnote{More recently, some subtleties associated with 
consistent gauge fixing world-volume diffeomorphisms and
$\kappa$-symmetry of superbrane actions have been discussed 
\cite{Pasti}. These subtleties do not change our conclusion.}
where the gauge condition
of the $\kappa$-symmetry was selected to be
%**   3.15 %%%%%%%%%%%%%%%%%%%%%%%%%%%%%%%%%%%%%%%%%%%%%%%%%%%%%%%%%
\begin{eqnarray}
\Theta^I_{-} = 0,
\label{3.15}
\end{eqnarray}
%%%%%%%%%%%%%%%%%%%%%%%%%%%%%%%%%%%%%%%%%%%%%%%%%%%%%%%%%%%%%%%%%%%
where 
%**   3.16 %%%%%%%%%%%%%%%%%%%%%%%%%%%%%%%%%%%%%%%%%%%%%%%%%%%%%%%%%
\begin{eqnarray}
\Theta^I_{\pm} \equiv  {\cal P}^{IJ}_{\pm} \Theta^J, \ 
{\cal P}^{IJ}_{\pm} \equiv \frac{1}{2}(\delta^{IJ}
\pm \Gamma_{0123} \epsilon^{IJ}).
\label{3.16}
\end{eqnarray}
%%%%%%%%%%%%%%%%%%%%%%%%%%%%%%%%%%%%%%%%%%%%%%%%%%%%%%%%%%%%%%%%%%%
One of the remakable things is that in this gauge (\ref{3.15})
the superfields take the simple forms given by
%**   3.17 %%%%%%%%%%%%%%%%%%%%%%%%%%%%%%%%%%%%%%%%%%%%%%%%%%%%%%%%%
\begin{eqnarray} 
L^t_{is} = \frac{1}{y} \partial_i y^t, \  L^I_{js} = s y^{\frac{1}{2}}
\partial_i \theta^I_{+},
\label{3.17}
\end{eqnarray}
%%%%%%%%%%%%%%%%%%%%%%%%%%%%%%%%%%%%%%%%%%%%%%%%%%%%%%%%%%%%%%%%%%%
where $t=4, \cdots, 9$, $y$ is a coordinate of $AdS_5 \times
S^5$ in the Cartesian coordinate, and $\Theta^I_{+} \equiv 
y^{\frac{1}{2}} \theta^I_{+}$ \cite{Kallosh1, Kallosh & Tseytlin}.

If we introduce the orthogonal matrix $U$ for the $SO(2)$
rotation of the spinor coordinates such that 
$\Theta^I = U^{IJ} \tilde{\Theta}^J$ where  
$U = \frac{1}{\sqrt{1 + (C_0 - \sqrt{1+C_0^2})^2}}[(- C_0 + 
\sqrt{1+C_0^2})1 - {\cal E}]$, it is easy to show that this
matrix $U$ indeed induces the $SO(2)$ rotation of the 
'Pauli matrices' (\ref{3.9}) since we have
%**   3.18 %%%%%%%%%%%%%%%%%%%%%%%%%%%%%%%%%%%%%%%%%%%%%%%%%%%%%%%%%
\begin{eqnarray}
U^T ({\cal I} - C_0 {\cal K}) U =  \sqrt{1 + C_0^2} {\cal K}.
\label{3.18}
\end{eqnarray}
%%%%%%%%%%%%%%%%%%%%%%%%%%%%%%%%%%%%%%%%%%%%%%%%%%%%%%%%%%%%%%%%%%%

As shown in the previous work \cite{Oda4}, the projector 
${\cal P}^{IJ}_{\pm}$ is invariant under the orthogonal
transformation by the matrix $U$, which yields the important
relation $\theta^I_{+} = U^{IJ} \tilde{\theta}^J_{+}$.
Since in the gauge condition (\ref{3.15}) the action of super
D3-brane becomes a very simple form, we can carry out
the transformation (\ref{3.18}) with keeping the other
irrelevant terms including the spinor coordinates unchanged.
(It is useful to use the expression (\ref{3.1}) rather than
(\ref{3.2}) of the action to prove this fact.)
Hence we have shown that the $SO(2)$ rotation of the 'Pauli
matrices' (\ref{3.9}) is mathematically allowed if we fix
the gauge associated with the $\kappa$-symmetry.

%%%%%%%%%%%%%%%%%%%%%%%%%%%%%%%%%%%%%%%%%%%%%%%%%%%%%%%%%%%%%%%%%%%%%
%%%%%%%%%%%%%%%%%%%%%%%%%%%%%%   SEC  4    %%%%%%%%%%%%%%%%%%%%%%%%%%
%%%%%%%%%%%%%%%%%%%%%%%%%%%%%%%%%%%%%%%%%%%%%%%%%%%%%%%%%%%%%%%%%%%%%
\section{Discussions}

In this paper, we have studied the property of the $SL(2,Z)$ 
self-duality of a supersymmetric and $\kappa$-symmetric D3-brane 
action in the $AdS_5 \times S^5$ background. We have seen that
on this background the dilaton and the axion undergo the same 
the $SL(2,Z)$ duality transformation as in the case of 
the flat Minkowskian background. 
It is interesting to notice that although the physical
contents of a theory in the two backgrounds are different, 
the derivation method of the $SL(2,Z)$ duality transformation 
is quite similar. We think that this similarity holds
even in proving the self-duality of super D-brane theories
in general background geometries. 

In this article, we have confined ourselves to a classical
analysis to prove the self-duality of the super D3-brane
action.
Of course, a quantum-mechanically exact analysis would
be welcoming in future.  However, this problem
may not be so important by the following two reasons.
One reason comes from the fact that the action of p-branes ($p \ge 2$) 
is in essense unrenormalizable so that new degrees of dynamical
freedom and new higher order terms in general appear in approaching 
the short distance regime. 
In this sense, the exact treatment within the present context,
i.e., showing the self-duality of classical action of super D3-brane
in a quantum-mechanical way, is not so interesting. 
The other stems from a reasoning somewhat contrary to
the first statement. Namely, the super D3-brane action on 
$AdS_5 \times S^5$ may be an exact conformal field theory and receive
no quantum corrections except the renormalization of an overall factor
owing to the no-go theorems
from the maximal supersymmetry. Actually we have already had some articles
discussing the related problem \cite{Banks, Rajaraman}. 
Accordingly, from this vantage point
the classical analysis presented in this article may be sufficient
as an exact proof.  

Finally we would like to make comments on future works.
One interesting direction is to construct general super
D-brane actions on $AdS_5 \times S^5$ and investigate
various duality transformations along the present analysis. 
In paticular, as understood
from an analysis in the flat background we know that super D2-brane 
and D4-brane actions would transform in a manner that is expected from 
the relation between type IIA superstring theory and 11 dimensional 
M theory. We believe that this behavior is also valid in the case of
the $AdS_5 \times S^5$. We would like to clarify this point
in near future. 

Another interesting problem which is also related to the first work
is to construct a M5-brane theory on $AdS_7 \times S^4$. 
The following strategy is perhaps useful,
namely, start with the supergroup $OSp(6,2|4)$ directly, consider the coset 
superspace $OSp(6,2|4)/(SO(6,1) \otimes SO(4))$, solve the Maurer-Cartan 
equations implied by the $osp(6,2|4)$ superalgebra, and then find the 
supersymmetric and $\kappa$-symmetric action, which would become the form 
as the PST action \cite{Mario} in the flat background limit except the 
Freund-Rubin coupling. 
We can conjecture that the D4-brane action would be also identical
to the action by double-dimensional reduction of the so-obtained
M5-brane action in the $AdS_7 \times S^4$ background as in the flat
background, so studies of the super D4-brane action may in turn
yield some useful informations in completing this program.
(See the related papers \cite{de Witt, Claus}.) 
We also return to this problem in future.

\vs 1
%%%%%%%%%%%%%%%%%%%%%%%%%%%%%%%%%%%%%%%%%%%%%%%%%%%%%%%%%%%%%%%%%%
%%%%%%%%%%%%%%%%%%%%%%%% Acknowledgement %%%%%%%%%%%%%%%%%%%%%%%%%%%%%
%%%%%%%%%%%%%%%%%%%%%%%%%%%%%%%%%%%%%%%%%%%%%%%%%%%%%%%%%%%%%%%%%%
\begin{flushleft}
{\bf Acknowledgement}
\end{flushleft}
This work was supported in part by Grant-Aid for Scientific 
Research from Ministry of Education, Science and Culture 
No.09740212.

\vs 1
%%%%%%%%%%%%%%%%%%%%%%%%%%%%%%%%%%%%%%%%%%%%%%%%%%%%%%%%%%%%%%%%
%%%%%%%%%%%%%%%%%%%%%%%% reference %%%%%%%%%%%%%%%%%%%%%%%%%%%%%%%
%%%%%%%%%%%%%%%%%%%%%%%%%%%%%%%%%%%%%%%%%%%%%%%%%%%%%%%%%%%%%%%%%%

\end{document}